\documentclass[12pt]{article}
\title{Darboux Covariant Equations of von~Neumann Type and their
Generalizations}
\author{Jan L.\ Cie\'sli\'nski$^1$, Marek Czachor$^{2,3}$ and
Nikolai~V.~Ustinov$^{4,2}$\\
\\
$^1$ Uniwersytet w Bia\l ymstoku, Instytut Fizyki Teoretycznej\\
ul.\ Lipowa 41, 15-424 Bia\l ystok, Poland\\
$^2$ Katedra Fizyki Teoretycznej i Metod Matematycznych\\
Politechnika Gda\'{n}ska, ul. Narutowicza 11/12, 80-952 Gda\'{n}sk, Poland\\
$^3$ Department of Physics\\
Technische Universit\"at Clausthal, 38678 Clausthal-Zellerfeld, Germany\\
$^4$ Theoretical Physics Department, Kaliningrad State University, \\
Al.\,Nevsky street 14, 236041, Kaliningrad, Russia\\
}
\date{ }
\begin{document}
\maketitle
\begin{abstract}
Lax pairs with operator valued coefficients, which are explicitly connected by 
means of an additional condition, are considered.
This condition is proved to be covariant with respect to the Darboux
transformation of a general form.
Nonlinear equations arising from the compatibility condition of the Lax pairs 
in the matrix case include, in particular, Nahm equations, Volterra, 
Bogoyavlenskii and Toda lattices.
The examples of another one--, two-- and multi--field lattice equations are 
also presented.
\end{abstract}

\pagebreak

\section{Introduction}

In the present paper we consider a large class of the integrable systems of 
nonlinear equations taking values on an associative ring of noncommutative 
operators. 
They are defined as the compatibility condition of Lax pairs characterized by 
the property that the equation for the time evolution of the wave function is 
explicitly determined (in a local way) by the coefficients of the spectral 
problem. 
There exist different types of Lax pairs, whose coefficients are connected 
explicitly.
Lax pairs with coefficients defined on an associative algebra of scalar
pseudo--differential operators were introduced in \cite{GD,K} (see also 
\cite{DS} and references therein).
These Lax pairs play an important role in the Sato theory \cite{Sato} and in 
constructing the modifications of KP hierarchy \cite{Cheng,SiSt}.
The case of the shift operators and associated lattice equations were 
discussed in \cite{B}.
The approach, which is applied below to connect the coefficients of the Lax 
pairs, differs from known ones.
The hierarchy of Darboux covariant nonlinear "multi--field" equations we 
describe in this paper contains, as the simplest case, the "one--field" 
equations of von~Neumann type
\begin{equation}
i\dot \rho=[H(\rho),\rho],\quad
i\dot \rho=[H,f(\rho)].
\label{vNe}
\end{equation}
These equations and their solutions were investigated in the context of the 
density matrices and Hamiltonians in \cite{LC98,KCzL,ULCK}, where, for 
instance, the formulas of the Darboux transformation were constructed. 
In the matrix case, the multi--field equations we derive here admit the 
reductions leading to known and new integrable nonlinear lattice systems.

The technique exploited in this paper combines and develops the approaches of 
works \cite{Ci,UC}.
In Sec.II we show that the relations between the coefficients of the equations 
forming the Lax pair, i.e. necessary conditions for the Lax pair 
compatibility, are identically satisfied if an additional condition on the 
coefficients is imposed.
This condition connects explicitly the coefficients of the equations of the
Lax pair and allows us to write in a closed form the nonlinear equations,
which follow the compatibility condition.
A theorem establishing the Darboux covariance of the Lax pair with the 
additional condition and, consequently, of the corresponding nonlinear 
equations is proved in Sec.III.
This theorem gives an effective tool of producing the infinite hierarchies of 
solutions, including the multi--soliton solutions, for nonlinear equations and 
their reductions.
The equations of von~Neumann type and associated lattice equations are 
considered in Sec.IV.
The next two sections are devoted to two-- and multi--field generalizations of
the von~Neumann type equations.
The examples presented include known lattice equations as well as some new 
ones.
Particular cases of the Darboux transformation satisfying the conditions of 
the theorem of Sec.III are discussed in the Appendix.

\section{Lax pair and multi--field equations}

Let us begin with the overdetermined system of linear equations (Lax pair) 
\begin{equation}
\left\{
\begin{array}{rcl}
-i\dot\psi&=&\psi A(\lambda)\\
z_{\lambda}\psi&=&\psi H(\lambda)
\end{array}
\right..
\label{Lax}
\end{equation}
Here $\lambda$ and $z_{\lambda}$ are complex numbers, $\psi$ takes values in a
given linear space $\cal L$, $A(\lambda)$ and $H(\lambda)$ are linear
operators $\cal L\mapsto\cal L$ belonging to an associative ring, the dot
denotes a derivative (i.e. an operator satisfying the Leibnitz rule).
The compatibility condition of Eqs.\ (\ref{Lax}) is
\begin{equation}
i\dot H(\lambda)=[A(\lambda),H(\lambda)].
\label{cc}
\end{equation}
If we assume that the operators entering the Lax pair are rational functions
of $\lambda$ with operator valued coefficients of the form
\begin{eqnarray}
A(\lambda)&=&\displaystyle
\sum_{k=0}^L\lambda^kB_k+\sum_{k=1}^M\frac{1}{\lambda^k}C_k,
\label{ABC}\\
H(\lambda)&=&\displaystyle \sum_{k=0}^N\lambda^kH_k.
\label{H}
\end{eqnarray}
then Eq.\ (\ref{cc}) becomes equivalent to the following system of algebraic
and differential relations between operators $B_k$, $C_k$ and $H_k$\,:
\begin{eqnarray}
\displaystyle\sum_{k=\max\{0,m-L\}_{\mathstrut}}^N[B_{m-k},H_k]&=&0
\qquad(N<m\le L+N),
\label{setB}\\
\displaystyle\sum_{k=0}^{\min\{N,m+M\}^{\mathstrut}}[C_{k-m},H_k]&=&0
\qquad(-M\le m<0),
\label{setC}
\end{eqnarray}
\begin{equation}
i\dot H_m=\sum_{k=\max\{0,m-L\}}^{m}[B_{m-k},H_k]+
\sum_{k=m+1}^{\min\{N,m+M\}}[C_{k-m},H_k]\qquad(0\le m\le N).
\label{ode}
\end{equation}

The connection between operators $B_k$, $C_k$ and $H_k$, which is implied by 
Eqs.\ (\ref{setB},\ref{setC}), is implicit.
It is possible to express $B_k$ and $C_k$ explicitly in terms of the operators 
$H_k$ in order to satisfy Eqs.\ (\ref{setB},\ref{setC}) {\it identically}.
Indeed, let us put
\begin{eqnarray}
B_k&=&\displaystyle
\frac{1}{(L-k)!}\left.\left(\frac{d^{L-k}}{d\varsigma^{L-k}}
f(\varsigma^NH(\varsigma^{-1}),\varsigma^{-1})\right)\right|_{\varsigma=0},
\label{B}\\
C_k&=&\displaystyle
\frac{1}{(M-k)!}\left.\left(\frac{d^{M-k}}{d\varepsilon^{M-k}}
g(H(\varepsilon),\varepsilon)\right)\right|_{\varepsilon=0},
\label{C}
\end{eqnarray}
where $f(X,\lambda)$ and $g(X,\lambda)$ are well defined functions of an
operator $X$ and parameter $\lambda$. 
We assume hereafter that for any operator $X(\lambda)$, which is analytic in 
the neighborhood of the point $\lambda=\infty$, the function 
$f(X(\lambda),\lambda)$ is also analytic and
the condition
\begin{equation}
[f(X(\lambda),\lambda),X(\lambda)]=0.
\label{f1}
\end{equation}
is valid in this neighborhood as well.
In the case of the function $g(X,\lambda)$, analogous properties have to take
place in the neighborhood of the point $\lambda=0$. 
In particular we have 
\begin{equation}
\label{g1}
[g(X(\lambda),\lambda),X(\lambda)]=0.
\end{equation}
Eqs.\ (\ref{setB},\ref{setC}) are fulfilled for any $B_k$ and $C_k$ defined by
Eqs.\ (\ref{B},\ref{C}) as a consequence of the following identities
$$
\frac{d^{N+L-m}}{d\varsigma^{N+L-m}}
[f(\varsigma^NH(\varsigma^{-1}),\varsigma^{-1}),\varsigma^NH(\varsigma^{-1})]
\Big|_{\varsigma=0}\equiv0\qquad(N<m\le L+N),
$$
$$
\frac{d^{M+m}}{d\varepsilon^{M+m}}
[g(H(\varepsilon),\varepsilon),H(\varepsilon)]\Big|_{\varepsilon=0}
\equiv0\qquad(-M\le m<0).
$$
Using Eqs.\ (\ref{setB},\ref{setC}) we can rewrite Eqs.\ (\ref{ode}) as
\begin{equation}
-i\dot H_m=\sum_{k=m+1}^N[B_{m-k},H_k]+\sum_{k=0}^m[C_{k-m},H_k],
\label{eode}
\end{equation}
where the coefficients $B_k$ and $C_k$ for $k<0$ are calculated accordingly to 
Eqs.\ (\ref{B},\ref{C}).
The operator $A(\lambda)$ given by (\ref{ABC},\ref{B},\ref{C}) is conveniently 
represented in the following equivalent way
\begin{equation}
A(\lambda)=[F(H(\lambda),\lambda)]_{\infty}+[G(H(\lambda),\lambda)]_{0}.
\label{AFG}
\end{equation}
Here
\begin{equation}
F(H(\lambda),\lambda)=\lambda^Lf(H(\lambda)/\lambda^N,\lambda),
\label{F}
\end{equation}
\begin{equation}
G(H(\lambda),\lambda)=\lambda^{-M}g(H(\lambda),\lambda),
\label{G}
\end{equation}
symbols $[...]_{\infty}$ and $[...]_{0}$ denote the parts of the power 
expansions in $\lambda$ that contain nonnegative and negative powers 
respectively.
These equations establish a connection between the coefficients of Lax pair 
(\ref{Lax}). 

In what follows we restrict our consideration to functions $f(X,\lambda)$ and
$g(X,\lambda)$, which possess an additional property, namely they are covariant
with respect to a similarity transformation $T$:
\begin{equation}
f(T^{-1}XT,\lambda)=T^{-1}f(X,\lambda)T,\qquad
g(T^{-1}XT,\lambda)=T^{-1}g(X,\lambda)T.
\label{fg2}
\end{equation}
Conditions (\ref{f1},\ref{g1},\ref{fg2}) are not very 
restrictive and are satisfied,
for example, by polynomials in $X$ and sums of negative powers of polynomials
in $X$ with scalar coefficients.
If $X$ is selfadjoint operator, then these conditions are valid for all 
$f(X,\lambda)$ and $g(X,\lambda)$ determined via the spectral theorem.

\section{Darboux covariance}

Let us consider the transformation
\begin{equation}
\psi[1]=\psi D(\lambda),
\label{Tpsi}
\end{equation}
where $...[1]$ denotes the image under the transformation, and $D(\lambda)$ is
an invertible linear operator depending on $\lambda$.
We say that the Lax pair (\ref{Lax}) is Darboux covariant with respect to
transformation
$\{\psi,A(\lambda),H(\lambda)\}\to\{\psi[1],A[1](\lambda),H[1](\lambda)\}$ if
the following equations hold
\begin{equation}
\left\{
\begin{array}{rcl}
-i\dot\psi[1]&=&\psi[1]A[1](\lambda)\\
z_{\lambda}\psi[1]&=&\psi[1]H[1](\lambda)
\end{array}
\right.
\label{TLax}
\end{equation}
and the structure of $A[1](\lambda)$ and $H[1](\lambda)$ is the same as
the structure of $A(\lambda)$ and $H(\lambda)$.
The notion of "structure" means that the shapes of the coefficients of Lax
pairs (\ref{Lax}) and (\ref{TLax}) are the same.
The most important point is that the locations of singularities of 
$A(\lambda)$ and $A[1](\lambda)$, $H(\lambda)$ and $H[1](\lambda)$ and their 
types should coincide.
The transformations of the form (\ref{Tpsi},\ref{TA},\ref{TH}) that satisfy 
these conditions are called Darboux transformations \cite{MatveevSalle}.
These transformations allow one to generate the hierarchies of solutions of
nonlinear equations admitting the compatibility condition representation and
of associated Lax pairs.
In finite dimensional (matrix) cases $D(\lambda)$ is termed the Darboux
matrix \cite{Ci1}.

Substituting (\ref{Tpsi}) into (\ref{TLax}) we obtain expressions for the
coefficients of the transformed Lax pair
\begin{equation}
A[1](\lambda)=-iD(\lambda)^{-1}\dot D(\lambda)+
D(\lambda)^{-1}A(\lambda)D(\lambda).
\label{TA}
\end{equation}
\begin{equation}
H[1](\lambda)=D(\lambda)^{-1}H(\lambda)D(\lambda).
\label{TH}
\end{equation}
If $D(\lambda)$ and $D(\lambda)^{-1}$ are regular on the plane of parameter 
$\lambda$ at singular points of the coefficients of Lax pair then the 
sufficient condition for $D(\lambda)$ to define the Darboux transformation 
comes from the requirement that the right hand sides in 
Eqs.\ (\ref{TA},\ref{TH}) have no the singularities at the points, which are 
the singular point of $D(\lambda)$ and $D^{-1}(\lambda)$.
We refer the Reader to the Appendix, where the examples of such
transformation are presented.

The following theorem gives sufficient conditions of covariance of 
Eq.\ (\ref{AFG}), which explicitly connects the coefficients of the Lax pair 
(\ref{Lax}), with respect to the Darboux transformation 
(\ref{Tpsi},\ref{TH},\ref{TH}).

{\bf Theorem.} {\it
If $D(\lambda)$ and $D(\lambda)^{-1}$ are rational functions in $\lambda$ that 
have poles at finite points $\mu_1,\ldots,\mu_s$ and $\mu_{s+1},\ldots,\mu_S$ respectively, $\mu_k\ne0$
($k=1,\ldots,S$) and $[D(\lambda)^{-1}\dot D(\lambda)]_{\infty}=0$, then
Eq.\ (\ref{AFG}) is Darboux covariant.}

{\it Proof}:
Substituting (\ref{AFG}) into (\ref{TA}) yields
\begin{equation}
A[1](\lambda)=-iD(\lambda)^{-1}\dot D+D(\lambda)^{-1}
[F(H(\lambda),\lambda)]_{\infty}D(\lambda)+
D(\lambda)^{-1}[G(H(\lambda),\lambda)]_{0}D(\lambda).
\label{TAFG}
\end{equation}
It is enough to show that
\begin{equation}
A[1](\lambda)=[F(H[1](\lambda),\lambda)]_{\infty}+
[G(H[1](\lambda),\lambda)]_{0}.
\label{TATFG}
\end{equation}

The right hand side of (\ref{TAFG}) is a rational function of $\lambda$ with
poles at most at $\mu_1,\ldots,\mu_S$ and $0$, $\infty$.
Therefore the following decomposition holds
\begin{equation}
A[1](\lambda)=\sum_{k=1}^{S}[A[1](\lambda)]_{\mu_k}+
[A[1](\lambda)]_{\infty}+[A[1](\lambda)]_{0}
\label{dTA}
\end{equation}
where $[A[1](\lambda)]_{\mu_k}$ is the principal part of the Laurent
expansion of $A[1](\lambda)$ at the point $\lambda=\mu_k$.
Since the first term in formula (\ref{TAFG}) can have poles at most at
$\mu_1,\ldots,\mu_S$ and vanishes as $\lambda\to\infty$, we obtain
$$
[A[1](\lambda)]_{\infty}=
[D(\lambda)^{-1}[F(H(\lambda),\lambda)]_{\infty}D(\lambda)]_{\infty}
+[D(\lambda)^{-1}[G(H(\lambda),\lambda)]_{0}D(\lambda)]_{\infty},
$$
$$
[A[1](\lambda)]_{0}=
[D(\lambda)^{-1}[F(H(\lambda),\lambda)]_{\infty}D(\lambda)]_{0}
+[D(\lambda)^{-1}[G(H(\lambda),\lambda)]_{0}D(\lambda)]_{0}.
$$
Using equalities
$$
\begin{array}{c}
[D(\lambda)^{-1}[F(H(\lambda),\lambda)]_{\infty}D(\lambda)]_{\infty}=
[D(\lambda)^{-1}F(H(\lambda),\lambda)D(\lambda)]_{\infty},
\\[3ex]
[D(\lambda)^{-1}[F(H(\lambda),\lambda)]_{\infty}D(\lambda)]_{0}=
[[D(\lambda)^{-1}F(H(\lambda),\lambda)D(\lambda)]_{\infty}]_{0}=0,
\\[3ex]
[D(\lambda)^{-1}[G(H(\lambda),\lambda)]_{0}D(\lambda)]_{\infty}=
[[D(\lambda)^{-1}G(H(\lambda),\lambda)D(\lambda)]_{0}]_{\infty}=0,
\\[3ex]
[D(\lambda)^{-1}[G(H(\lambda),\lambda)]_{0}D(\lambda)]_{0}=
[D(\lambda)^{-1}G(H(\lambda),\lambda)D(\lambda)]_{0}.
\end{array}
$$
we have
\begin{eqnarray}
[A[1](\lambda)]_{\infty}=
[D(\lambda)^{-1}F(H(\lambda),\lambda)D(\lambda)]_{\infty},
\\[3ex]
[A[1](\lambda)]_{0}=[D(\lambda)^{-1}G(H(\lambda),\lambda)D(\lambda)]_{0}.
\label{dTApn}
\end{eqnarray}
It follows from the definition of the Darboux transformation
that for $k=1,\ldots,S$
\begin{equation}
[A[1](\lambda)]_{\mu_k}\equiv0 \ .
\label{mu_k}
\end{equation}
Combining (\ref{dTA}--\ref{mu_k}) and taking into account (\ref{TH}), 
(\ref{F}--\ref{fg2}), we get (\ref{TATFG}).
\rule{5pt}{5pt}

It is well known that the conditions (\ref{mu_k}) can be solved resulting in
an explicit expression for $D(\lambda)$ in terms of solutions of the Lax
pair (\ref{Lax}) and a dual pair, which belong to the kernel of this operator
or its inverse one.
For this reason, we prefer to use the name of the Darboux transformation 
technique instead of the dressing method. 
In the Appendix we give examples of the Darboux transformations, which
satisfy the conditions of our Theorem and can be used to construct the
hierarchies of solutions of nonlinear equations (\ref{ode}) under constraints
(\ref{B},\ref{C}).
The applications of the Darboux transformation technique to certain nonlinear 
equations of von~Neumann type, including an equation in infinite dimensional
case, which are of interest in connection with quantum mechanics and
statistical physics, can be found in \cite{LC98,KCzL,ULCK}.

If operator $T$ in Eqs.\ (\ref{fg2}) is independent of $\lambda$, then the 
expressions in right hand sides of these formulas correspond to so-called 
gauge transformation of wave function 
$$
\psi\to\tilde\psi=\psi\,T.
$$
The case of the Lax pairs with $g(X,\lambda)\equiv0$ is gauge equivalent to 
the case $f(X,\lambda)\equiv0$ if $T$ solves equation
$$
i\dot T=B_0T.
$$
Some of the lattice equations presented in the next sections are the 
compatibility condition of gauge equivalent Lax pairs.

\section{Equations of von~Neumann type (one--field equations)}

In this section we consider the case $N=1$, i.e.,
$$
H(\lambda)=\lambda H_1+H_0.
$$
The compatibility condition of the Lax pair gives us the equation 
\begin{equation}
i\dot H_0=[B_0,H_0]+[C_1,H_1],
\label{vNt}
\end{equation}
$$
\dot H_1=0,
$$
where
\begin{equation}
B_0=\frac{1}{L!}\left.\left(\frac{d^L}{d\varsigma^L}
f(H_1+\varsigma H_0),\varsigma^{-1})\right)\right|_{\varsigma=0},
\label{B_0}
\end{equation}
\begin{equation}
C_1=\frac{1}{(M-1)!}\left.\left(\frac{d^{M-1}}{d\varepsilon^{M-1}}
g(\varepsilon H_1+H_0,\varepsilon)\right)\right|_{\varepsilon=0}.
\label{C_1}
\end{equation}
We refer to this case as "one--field" because $H_1$ is the constant operator. 
The right--hand side of Eq.\ (\ref{vNt}) combines both types of the 
nonlinearities as in the equations of von~Neumann type (\ref{vNe}).
The nonlinear equations corresponding to simplest choices of functions
$f(X,\lambda)$ and $g(X,\lambda)$ are presented below.
We also obtain the equations that follow them if the matrix coefficients $H_1$ 
and $H_0$ are defined in a special manner.

\subsection{$f(X,\lambda)=iX^l$ ($l\in\bf N$), $g(X,\lambda)=0$, $L=1$}
%1

The compatibility condition leads to the equation
\begin{equation}
\dot H_0=\sum_{m=1}^l[H_1^{m-1}H_0H_1^{l-m},H_0].
\label{cc11}
\end{equation}
This equation with $l=2$ is multi--dimensional Euler's top equation
\cite{A,Mishch,M1}.
Darboux covariance of Eq.\ (\ref{cc11}) and associated Lax pair was proved in
\cite{KCzL}.

Let matrices $H_1$ and $H_0$ have the form
\begin{equation}
H_{1,kj}=\delta_{k,j-1},\
H_{0,kj}=\rho_k\delta_{k,j+l-1}.
\label{H01_1}
\end{equation}
Then Eq.\ (\ref{cc11}) yields
\begin{equation}
\dot\rho_k=\rho_k\displaystyle\sum_{m=1}^{l-1}(\rho_{k+m}-\rho_{k-m}).
\label{Bl}
\end{equation}
These equations are known as the Bogoyavlenskii lattice \cite{B1}.
In the case $l=2$ Eqs.\ (\ref{Bl}) coincide with the Volterra system \cite{V}
\begin{equation}
\dot\rho_k=\rho_k(\rho_{k+1}-\rho_{k-1}),
\label{V}
\end{equation}
which describes stimulated scattering of plasma oscillations by ions
\cite{ZMR}.

\subsection{$f(X,\lambda)=iX^{-n}$ ($n\in\bf N$), $g(X,\lambda)=0$, $L=1$}
%2

In this case Eq.\ (\ref{vNt}) is written as given
\begin{equation}
\dot H_0=-\sum_{m=1}^n[H_1^{-m}H_0H_1^{m-n-1},H_0].
\label{cc12}
\end{equation}
The Bogoyavlenskii lattice (\ref{Bl}) with $l=n+1$ follows this equation if 
matrices $H_1$ and $H_0$ are chosen in the form
\begin{equation}
H_{1,kj}=\delta_{k,j-1},\
H_{0,kj}=\rho_k\delta_{k,j-n-1}.
\label{H01_2}
\end{equation}

\subsection{$f(X,\lambda)=0$, $g(X,\lambda)=iX^l$ ($l\in\bf N$), $M=1$}
%3

From Eq.\ (\ref{vNt}) we have
\begin{equation}
\dot H_0=[H_0^l,H_1].
\label{cc13}
\end{equation}
Assuming that matrices $H_1$ and $H_0$ are represented in the following manner
\begin{equation}
H_{1,kj}=\delta_{k,j-l+1},\
H_{0,kj}=\rho_k\delta_{k,j+1},
\label{H01_3}
\end{equation}
we obtain the well known lattice \cite{B1}
\begin{equation}
\dot\rho_k=\displaystyle\prod_{m=0}^{l-1}\rho_{k-m}-
\prod_{m=0}^{l-1}\rho_{k+m}.
\label{r_3}
\end{equation}
These equations with $l=2$ are obviously reduced to the Volterra system 
(\ref{V}).
If we put
$$
\rho_k=\exp ({\displaystyle u_k} ) \ ,
$$
then Eqs.\ (\ref{r_3}) read as 
\begin{equation}
\dot u_k= \exp ({\sum_{m=1}^{l-1}\displaystyle u_{k-m}})-
\exp ({\sum_{m=1}^{l-1}\displaystyle u_{k+m}}) .
\label{u_3}
\end{equation}
As it was noted at the end of Sec.III this case is gauge equivalent to the 
case IV.1.

\subsection{$f(X,\lambda)=0$, $g(X,\lambda)=iX^l$ ($l\in\bf N$), $M=2$}
%4

Eq.\ (\ref{vNt}) yields
\begin{equation}
\dot H_0=\sum_{m=0}^{l-1}[H_0^mH_1H_0^{l-m-1},H_1].
\label{cc14}
\end{equation}
Supposing
\begin{equation}
H_{1,kj}=\delta_{k,j-l+2},\
H_{0,kj}=\rho_k\delta_{k,j+2},
\label{H01_4}
\end{equation}
we have
\begin{equation}
\dot\rho_k=\displaystyle\sum_{m=0}^{l-1}\Bigl(\,
\prod_{i=0}^{m-1}\rho_{k-2i}\prod_{i=0}^{l-m-2}\rho_{k+2i-l+2}-
\prod_{i=0}^{m-1}\rho_{k-2i+l-2}\prod_{i=0}^{l-m-2}\rho_{k+2i}
\Bigr).
\label{r_4}
\end{equation}
(It is assumed hereafter that $\prod_{i=0}^{m}{...}_{\,i}=1$ if
$m<0$.)
In the case $l=3$ these equations are equivalent to the Volterra system
(\ref{V}).

\subsection{$f(X,\lambda)=0$, $g(X,\lambda)=iX^{-l}$ ($l\in\bf N$), $M=1$}
%5

Eq.\ (\ref{vNt}) takes the form
\begin{equation}
\dot H_0=[H_0^{-l},H_1].
\label{cc15}
\end{equation}
If matrices $H_1$ and $H_0$ are defined as follows
\begin{equation}
H_{1,kj}=\delta_{k,j+l+1},\
H_{0,kj}=\rho_k\delta_{k,j+1},
\label{H01_5}
\end{equation}
then the compatibility condition implies
\begin{equation}
\dot\rho_k=\displaystyle\prod_{m=1}^l\rho_{k+m}^{-1}-
\prod_{m=1}^l\rho_{k-m}^{-1}.
\label{r_5}
\end{equation}
Introducing two sets of new dependent variables
$$
\rho_k=\exp ({\displaystyle-u_k}),
$$
$$
v_k=\rho_k^{-1},
$$
we obtain equivalent representations of Eqs.\ (\ref{r_5})
\begin{equation}
\dot u_k=\exp ( \sum_{m=0}^l\displaystyle u_{k-m} ) -
\exp ( \sum_{m=0}^l\displaystyle u_{k+m} ) 
\label{u_5}
\end{equation}
and
\begin{equation}
\dot v_k=\displaystyle v_k^2\Bigl(\,\prod_{m=1}^lv_{k-m}-
\prod_{m=1}^lv_{k+m}\Bigr).
\label{v_5}
\end{equation}
Eqs.\ (\ref{u_5}) with $l=2$ were studied in \cite{HW}.
Lax pair for Eqs.\ (\ref{v_5}) was found in \cite{B1}.
These equations in the case $l=1$ obey a symmetry $v_k\to-v_k$ and look like a 
natural generalization of the Volterra system (\ref{V}).

\subsection{$f(X,\lambda)=0$, $g(X,\lambda)=iX^{-l}$ ($l\in\bf N$), $M=2$}
%6

In this case Eq.\ (\ref{vNt}) is written in the next manner 
\begin{equation}
\dot H_0=-\sum_{m=1}^l[H_0^{-m}H_1H_0^{m-l-1},H_1].
\label{cc16}
\end{equation}
If matrices $H_1$ and $H_0$ are defined as follows
\begin{equation}
H_{1,kj}=\delta_{k,j+l+2}, \qquad
H_{0,kj}=\rho_k\delta_{k,j+2},
\label{H01_6}
\end{equation}
then we come to equations
\begin{equation}
\dot\rho_k=\displaystyle\sum_{m=1}^l\Bigl(\,
\prod_{i=1}^m\rho_{k+2i-l-2}^{-1}\prod_{i=0}^{l-m}\rho_{k-2i-2}^{-1}-
\prod_{i=1}^m\rho_{k+2i}^{-1}\prod_{i=0}^{l-m}\rho_{k-2i+l}^{-1}
\Bigr).
\label{r_6}
\end{equation}

\subsection{$f(X,\lambda)=0$, $g(X,\lambda)=g(X)$, $M=1$}
%7

Here we have
\begin{equation}
i\dot H_0=[g(H_0),H_1].
\label{cc17}
\end{equation}
By construction, $g(H_0)$ commutes with $H_0$.
The Lax representation and Darboux covariance properties of Eq.\ (\ref{cc17}) 
with arbitrary well--defined function $g(X)$ were established in \cite{ULCK}.
The cases $g(X)=iX^3$ and $g(X)=iX^{-1}$ were considered in \cite{MS} in the
framework of the symmetry approach to the classification problem of integrable 
equations on free associative rings.
The Lax representation for the equations in these cases seems to be new.

\subsection{$f(X,\lambda)=X^4$, $g(X,\lambda)=0$, $L=2$}

The compatibility condition (\ref{vNt}) becomes
\begin{equation}
i\dot H_0=[h(H_0),H_0]=[H_1,F(H_0)]
\label{cc1}
\end{equation}
(compare with Eqs.\ (\ref{vNe})), where
$$
h(H_0)=H_0^2H_1^2+H_0H_1H_0H_1+H_0H_1^2H_0+H_1H_0^2H_1+H_1H_0H_1H_0+H_1^2H_0^2,
$$
$$
F(H_0)=H_0^3H_1+H_0^2H_1H_0+H_0H_1H_0^2+H_1H_0^3.
$$
Let us note that, contrary to the previous example, $[F(H_0),H_0]\neq0$.
We refer to the maps $H_0\mapsto F(H_0)$ of such a kind as {\it nonabelian 
functions\/}, or {\it nonabelian nonlinearities\/} \cite{UC}.
This example is a particular case of the equations (\ref{cc14}).

\section{Two--Field Equations}

A few examples of systems appearing if $N=2$ are considered in this section.

\subsection{$f(X,\lambda)=iX$, $g(X,\lambda)=0$, $L=1$}
%8

The compatibility condition (\ref{ode}) leads to equations
\begin{eqnarray}
\dot H_0&=&[H_1,H_0],\nonumber\\
\dot H_1&=&[H_2,H_0],\nonumber\\
\dot H_2&=&0.\nonumber
\end{eqnarray}
It is checked immediately that functions
\begin{eqnarray}
F_1&=&(H_2-\sigma H_0)/(2i),\nonumber\\
F_2&=&(H_2+\sigma H_0)/2,\nonumber\\
F_3&=&H_1/(2i),\nonumber
\end{eqnarray}
where $\sigma$ is a parameter ($\sigma\ne0$), satisfy equations
\begin{eqnarray}
\dot F_1&=&[F_2,F_3]+i[F_3,F_1],
\label{NF_1}\\
\dot F_2&=&[F_3,F_1]+i[F_3,F_2],
\label{NF_2}\\
\dot F_3&=&\sigma[F_1,F_2].
\label{NF_3}
\end{eqnarray}
In terms of new dependent variables
$$
f_k=g F_kg^{-1}\quad(k=1,2,3),
$$
where $g$ solves equation
$$
\dot g=-igF_3,
$$
Eqs.\ (\ref{NF_1}--\ref{NF_3}) are rewritten as
\begin{eqnarray}
\dot f_1&=&[f_2,f_3],\nonumber\\
\dot f_2&=&[f_3,f_1],\nonumber\\
\dot f_3&=&\sigma[f_1,f_2].\nonumber
\end{eqnarray}
If we impose condition $f_k^+=-f_k$, then $\sigma$ has to be real.
This system with $\sigma=1$ is known as Nahm equations \cite{Nahm,Hitchin}.
It will be shown in the next subsection that this case is also connected with
Toda lattice equation.

\subsection{$f(X,\lambda)=iX^l$ ($l\in\bf N$), $g(X,\lambda)=0$, $L=1$}
%9

From Eqs.\ (\ref{ode}) we have
\begin{eqnarray}
\dot H_0&=&\sum_{m=1}^l[H_2^{m-1}H_1H_2^{l-m},H_0],
\label{H_02}\\
\dot H_1&=&[H_2^l,H_0]+\sum_{m=1}^l[H_2^{m-1}H_1H_2^{l-m},H_1],
\label{H_12}\\
\dot H_2&=&0.\nonumber
\end{eqnarray}
If we put
\begin{equation}
H_{2,kj}=\delta_{k,j-1}, \qquad
H_{1,kj}=h_k\delta_{k,j+l-1}, \qquad
H_{0,kj}=\rho_k\delta_{k,j+2l-1},
\label{H012_2}
\end{equation}
then Eqs.\ (\ref{H_02},\ref{H_12}) read as
\begin{equation}
\left\{
\begin{array}{l}
\dot h_k=\rho_{k+l}-\rho_k+
h_k\displaystyle\sum_{m=1}^{l-1}(h_{k+m}-h_{k-m}),\\
\dot\rho_k=\rho_k\displaystyle\sum_{m=0}^{l-1}(h_{k+m}-h_{k-l-m}).
\end{array}
\right.
\label{hr_2}
\end{equation}
Let
$$
h_k=\dot\sigma_k.
$$
If coefficients $\rho_k$ are chosen as given 
$$
\rho_k=C\mbox{e}^{\sum_{m=0}^{l-1}\displaystyle(\sigma_{k+m}-\sigma_{k-l-m})}
$$
($C$ is arbitrary constant), then system (\ref{hr_2}) is equivalent to the
following equations
\begin{equation}
\ddot\sigma_k=C\Bigl(
\mbox{e}^{\sum_{m=0}^{l-1}\displaystyle(\sigma_{k+m+l}-\sigma_{k-m})}
-\mbox{e}^{\sum_{m=0}^{l-1}\displaystyle(\sigma_{k+m}-\sigma_{k-l-m})}
\Bigr)
+\dot\sigma_k\displaystyle\sum_{m=1}^{l-1}(\dot\sigma_{k+m}-\dot\sigma_{k-m}).
\label{s_2}
\end{equation}
Assuming $l=1$ and $C=1$ we come to the Toda lattice equation
\cite{T1,M2,F}
\begin{equation}
\ddot\sigma_k=\mbox{e}^{\displaystyle\sigma_{k+1}-\sigma_k}
-\mbox{e}^{\displaystyle\sigma_k-\sigma_{k-1}}.
\label{Tl}
\end{equation}
Eqs.\ (\ref{s_2}) can be viewed as a generalization of the Toda lattice on
the systems of particles interacting with 
finite number of nearest neighborhoods.
For $n\times n$ matrices, these equations admit additional reductions
$$
\sigma_{m+1+k}=-\sigma_{m+1-k} \quad \mbox{ if } \quad n=2m+1,
$$
$$
\sigma_{m+k}=-\sigma_{m+1-k} \quad \mbox{ if } \quad n=2m
$$
or
$$
\sigma_{m+k}=-\sigma_{m-k} \quad \mbox{ if } \quad n=2m
$$
($k=0,...,m$).
In the case $l=1$, these reductions lead to generalized periodic Toda
lattices, whose connection with the root systems of semisimple Lie algebras
was established in \cite{B2}.

\subsection{$f(X,\lambda)=iX^{-l}$ ($l\in\bf N$), $g(X,\lambda)=0$, $L=1$}
%10

The compatibility condition in this case yields
\begin{eqnarray}
\dot H_0&=&-\sum_{m=1}^l[H_2^{-m}H_1H_2^{m-l-1},H_0],
\label{H_03}\\
\dot H_1&=&[H_2^{-l},H_0]-\sum_{m=1}^l[H_2^{-m}H_1H_2^{m-l-1},H_1],
\label{H_13}\\
\dot H_2&=&0.\nonumber
\end{eqnarray}
If $H_2$, $H_1$ and $H_0$ are defined in the following manner
\begin{equation}
H_{2,kj}=\delta_{k,j+1},\qquad
H_{1,kj}=h_k\delta_{k,j+l+1},\qquad
H_{0,kj}=\rho_k\delta_{k,j+2l+1},
\label{H012_3}
\end{equation}
then Eqs.\ (\ref{H_03},\ref{H_13}) are written as
\begin{equation}
\left\{
\begin{array}{l}
\dot h_k=\rho_{k+l}-\rho_k-
h_k\displaystyle\sum_{m=1}^l(h_{k+m}-h_{k-m}),\\
\dot\rho_k=\rho_k\displaystyle\sum_{m=1}^l(h_{k-l-m}-h_{k+m}).
\end{array}
\right.
\label{hr_3}
\end{equation}
In the case $l=1$ these equations are so-called Belov--Chaltikian lattice
\cite{BCh}.
The bilinear approach was applied to Belov--Chaltikian lattice in \cite{HT1}.

Expressing dependent variables in the terms of new ones 
$$
h_k=\dot\sigma_k,
$$
$$
\rho_k=C\mbox{e}^{\sum_{m=1}^l\displaystyle(\sigma_{k-l-m}-\sigma_{k+m})}
$$
($C$ is a constant) we reduce Eqs.\ (\ref{hr_3}) to the Toda--type lattice 
equations 
\begin{equation}
\ddot\sigma_k=C\Bigl(
\mbox{e}^{\sum_{m=1}^l\displaystyle(\sigma_{k-m}-\sigma_{k+l+m})}
-\mbox{e}^{\sum_{m=1}^l\displaystyle(\sigma_{k-l-m}-\sigma_{k+m})}
\Bigr)
-\dot\sigma_k\displaystyle\sum_{m=1}^l(\dot\sigma_{k+m}-\dot\sigma_{k-m}).
\label{s_3}
\end{equation}

\subsection{$f(X,\lambda)=0$, $g(X,\lambda)=iX^l$ ($l\in\bf N$), $M=1$}
%11

In this case Eqs.\ (\ref{ode}) are rewritten as given 
\begin{eqnarray}
\dot H_0&=&[H_0^l,H_1],
\label{H_04}\\
\dot H_1&=&[H_0^l,H_2],
\label{H_14}\\
\dot H_2&=&0.\nonumber
\end{eqnarray}
If matrices $H_2$, $H_1$ and $H_0$ have the form
\begin{equation}
H_{2,kj}=\delta_{k,j-2l+1},\qquad
H_{1,kj}=h_k\delta_{k,j-l+1},\qquad
H_{0,kj}=\rho_k\delta_{k,j+1},
\label{H012_4}
\end{equation}
then Eqs.\ (\ref{H_04},\ref{H_14}) yield
\begin{equation}
\left\{
\begin{array}{l}
\dot h_k=\displaystyle\prod_{m=0}^{l-1}\rho_{k-m}-
\prod_{m=0}^{l-1}\rho_{k+l+m},\\
\dot\rho_k=\displaystyle h_{k-l}\prod_{m=0}^{l-1}\rho_{k-m}-
h_k\prod_{m=0}^{l-1}\rho_{k+m}.
\end{array}
\right.
\label{hr_4}
\end{equation}
These equations with $l=1$ are equivalent to the Toda lattice (\ref{Tl}).

\subsection{$f(X,\lambda)=0$, $g(X,\lambda)=iX^l$ ($l\in\bf N$), $M=2$}
%12

From Eqs.\ (\ref{ode}) we have
\begin{eqnarray}
\dot H_0&=&[H_0^l,H_2]+\sum_{m=0}^{l-1}[H_0^mH_1H_0^{l-m-1},H_1],
\label{H_05}\\
\dot H_1&=&\sum_{m=0}^{l-1}[H_0^mH_1H_0^{l-m-1},H_2],
\label{H_15}\\
\dot H_2&=&0.\nonumber
\end{eqnarray}
Taking matrices $H_2$, $H_1$ and $H_0$ as given
\begin{equation}
H_{2,kj}=\delta_{k,j-2l+2},\qquad
H_{1,kj}=h_k\delta_{k,j-l+2},\qquad
H_{0,kj}=\rho_k\delta_{k,j+2},
\label{H012_5}
\end{equation}
we put compatibility condition into the form
\begin{equation}
\left\{
\begin{array}{l}
\dot h_k=\displaystyle\sum_{m=0}^{l-1}\Bigl(\,
h_{k-2m}\prod_{i=0}^{m-1}\rho_{k-2i}\prod_{i=0}^{l-m-2}\rho_{k+2i-l+2}-
h_{k-2m+l-2}\prod_{i=0}^{m-1}\rho_{k-2i+l-2}\prod_{i=0}^{l-m-2}\rho_{k+2i}
\Bigr),\\
\dot\rho_k=h_{k-l}\displaystyle\sum_{m=0}^{l-1}h_{k-2m}
\prod_{i=0}^{m-1}\rho_{k-2i}\prod_{i=0}^{l-m-2}\rho_{k+2i-l+2}-{}\\
\displaystyle{}-h_k\sum_{m=0}^{l-1}h_{k-2m-2}
\prod_{i=0}^{m-1}\rho_{k-2i-2}\prod_{i=0}^{l-m-2}\rho_{k+2i-l}
+\prod_{i=0}^{l-1}\rho_{k-2i}-\prod_{i=0}^{l-1}\rho_{k+2i}.
\end{array}
\right.
\label{hr_5}
\end{equation}
In the case $l=2$ and $h_k=0$ these equations coincide with the
Volterra system (\ref{V}).

\subsection{$f(X,\lambda)=0$, $g(X,\lambda)=iX^{-l}$ ($l\in\bf N$), $M=1$}
%13

In this case Eqs.\ (\ref{ode}) yield
\begin{eqnarray}
\dot H_0&=&[H_0^{-l},H_1],
\label{H_06}\\
\dot H_1&=&[H_0^{-l},H_2],
\label{H_16}\\
\dot H_2&=&0.\nonumber
\end{eqnarray}
Let the matrices $H_2$, $H_1$ and $H_0$ be represented in the following manner
\begin{equation}
H_{2,kj}=\delta_{k,j+2l+1},\qquad
H_{1,kj}=h_k\delta_{k,j+l+1},\qquad
H_{0,kj}=\rho_k\delta_{k,j+1}.
\label{H012_6}
\end{equation}
The compatibility condition leads to the system
\begin{equation}
\left\{
\begin{array}{l}
\dot h_k=\displaystyle\prod_{m=1}^l\rho_{k+m}^{-l}-
\prod_{m=1}^l\rho_{k-l-m}^{-l},\\
\dot\rho_k=\displaystyle h_{k+l}\prod_{m=1}^l\rho_{k+m}^{-l}-
h_k\prod_{m=1}^l\rho_{k-m}^{-l}.
\end{array}
\right.
\label{hr_6}
\end{equation}

\subsection{$f(X,\lambda)=0$, $g(X,\lambda)=iX^{-l}$ ($l\in\bf N$), $M=2$}
%14

Eqs.\ (\ref{ode}) give
\begin{eqnarray}
\dot H_0&=&[H_0^{-l},H_2]-\sum_{m=1}^l[H_0^{-m}H_1H_0^{m-l-1},H_1],
\label{H_05a}\\
\dot H_1&=&-\sum_{m=1}^l[H_0^{-m}H_1H_0^{m-l-1},H_2],
\label{H_15a}\\
\dot H_2&=&0.\nonumber
\end{eqnarray}
Supposing
\begin{equation}
H_{2,kj}=\delta_{k,j+2l+2},\qquad
H_{1,kj}=h_k\delta_{k,j+l+2},\qquad
H_{0,kj}=\rho_k\delta_{k,j+2},
\label{H012_5a}
\end{equation}
we have 
\begin{equation}
\left\{
\begin{array}{l}
\dot h_k=\displaystyle\sum_{m=1}^l\Bigl(\,
h_{k+2m-2l-2}\prod_{i=1}^m\rho^{-1}_{k+2i-2l-2}
\prod_{i=0}^{l-m}\rho^{-1}_{k-2i-l-2}-
h_{k+2m}\prod_{i=1}^m\rho^{-1}_{k+2i}
\prod_{i=0}^{l-m}\rho^{-1}_{k-2i+l}
\Bigr),\\
\dot\rho_k=h_k\displaystyle\sum_{m=1}^lh_{k+2m-l-2}
\prod_{i=1}^m\rho^{-1}_{k+2i-l-2}\prod_{i=0}^{l-m}\rho^{-1}_{k-2i-2}-{}\\
\displaystyle{}-h_{k+l}\sum_{m=1}^lh_{k+2m}
\prod_{i=1}^m\rho^{-1}_{k+2i}\prod_{i=0}^{l-m}\rho^{-1}_{k-2i+l}
+\prod_{i=1}^l\rho^{-1}_{k+2i}-\prod_{i=1}^l\rho^{-1}_{k-2i}.
\end{array}
\right.
\label{hr_5a}
\end{equation}

\subsection{$f(X,\lambda)=iX^2$, $g(X,\lambda)=0$, $L=2$}
%15

The compatibility condition (\ref{ode}) in this case is written as
\begin{eqnarray}
\dot H_0&=&[H_1^2,H_0]+[H_2,H_0^2],\nonumber\\
\dot H_1&=&[H_2,H_0H_1+H_1H_0],\nonumber\\
\dot H_2&=&0.\nonumber
\end{eqnarray}
Taking matrices $H_2$, $H_1$ and $H_0$ as follows
\begin{equation}
H_{2,kj}=\delta_{k,j-1},\qquad
H_{1,kj}=h_k\delta_{k,j},\qquad
H_{0,kj}=\rho_k\delta_{k,j+1},
\label{H012_7}
\end{equation}
we obtain
\begin{equation}
\left\{
\begin{array}{l}
\dot h_k=\rho_{k+1}(h_{k+1}+h_k)-\rho_k(h_k+h_{k-1}),\\
\dot\rho_k=\rho_k(\rho_{k+1}-\rho_{k-1}+h_k^2-h_{k-1}^2).
\end{array}
\right.
\label{hr_7}
\end{equation}
This system is the first member of the hierarchy of higher Toda lattices 
\cite{T2,Per}.
In the case $h_k=0$ this system is nothing but the Volterra system (\ref{V}).

\section{Multi--Field Equations}

In this section we present nonlinear equations that follow Eq.\ (\ref{ode}) 
with positive integer $N$ for special choices of functions $f(X,\lambda)$ and 
$g(X,\lambda)$.

\subsection{$f(X,\lambda)=iX^l$ ($l\in\bf N$), $g(X,\lambda)=0$, $L=1$}
%16

The compatibility condition is 
\begin{equation}
\dot H_i=[H_N^l,H_{i-1}]+\sum_{m=1}^l[H_N^{m-1}H_{N-1}N_N^{l-m},H_i]
\label{ccN_1}
\end{equation}
($i=0,...,N$).
If we put
\begin{equation}
H_{i,kj}=\rho_k^{(i)}\delta_{k,j+(N-i)l-1}
\label{HN_1}
\end{equation}
($\rho_k^{(N)}=1$),
then Eqs.\ (\ref{ccN_1}) give
\begin{equation}
\dot\rho_k^{(i)}=\rho_{k+l}^{(i-1)}-\rho_k^{(i-1)}+\rho_k^{(i)}\displaystyle
\sum_{m=0}^{l-1}(\rho_{k+m}^{(N-1)}-\rho_{k+(i-N+1)l-m}^{(N-1)}).
\label{rN_1}
\end{equation}

\subsection{$f(X,\lambda)=iX^{-l}$ ($l\in\bf N$), $g(X,\lambda)=0$, $L=1$}
%17

In this case Eqs.\ (\ref{ode}) are written in the following way
\begin{equation}
\dot H_i=[H_N^{-l},H_{i-1}]-\sum_{m=1}^l[H_N^{-m}H_{N-1}N_N^{m-l-1},H_i]
\label{ccN_2}
\end{equation}
($i=0,...,N$).
Assuming
\begin{equation}
H_{i,kj}=\rho_k^{(i)}\delta_{k,j-(N-i)l-1},\qquad\rho_k^{(N)}=1,
\label{HN_2}
\end{equation}
we obtain
\begin{equation}
\dot\rho_k^{(i)}=\rho_{k-l}^{(i-1)}-\rho_k^{(i-1)}+\rho_k^{(i)}\displaystyle
\sum_{m=1}^l(\rho_{k+(N-i-1)l+m}^{(N-1)}-\rho_{k-m}^{(N-1)}).
\label{rN_2}
\end{equation}

\subsection{$f(X,\lambda)=0$, $g(X,\lambda)=iX^l$ ($l\in\bf N$), $M=1$}
%18

The compatibility condition (\ref{ode}) yield 
\begin{equation}
\dot H_i=[H_0^l,H_{i+1}]
\label{ccN_3}
\end{equation}
($i=0,...,N$).
Let matrices $H_i$ have the form
\begin{equation}
H_{i,kj}=\rho_k^{(i)}\delta_{k,j-il+1},
\label{HN_3}
\end{equation}
where $\rho_k^{(N)}=1$.
In this case Eqs.\ (\ref{ccN_3}) lead to the following lattice equations
\begin{equation}
\dot\rho_k^{(i)}=\displaystyle
\rho_{k-l}^{(i+1)}\prod_{m=0}^{l-1}\rho_{k-m}^{(0)}-
\rho_k^{(i+1)}\prod_{m=0}^{l-1}\rho_{k+il+m}^{(0)}.
\label{rN_3}
\end{equation}
The case $l=1$ was studied in \cite{BM,HT2}

\subsection{$f(X,\lambda)=0$, $g(X,\lambda)=iX^{-l}$ ($l\in\bf N$), $M=1$}
%19

From Eqs.\ (\ref{ode}) we have
\begin{equation}
\dot H_i=[H_0^{-l},H_{i+1}]
\label{ccN_4}
\end{equation}
($i=0,...,N$).
If matrices $H_i$ are defined as follows
\begin{equation}
H_{i,kj}=\rho_k^{(i)}\delta_{k,j+il+1},
\label{HN_4}
\end{equation}
then Eqs.\ (\ref{ccN_4}) give 
\begin{equation}
\dot\rho_k^{(i)}=\displaystyle
\rho_{k+l}^{(i+1)}\prod_{m=1}^l\rho_{k+m}^{-1}-
\rho_k^{(i+1)}\prod_{m=1}^l\rho_{k-il-m}^{-1},
\label{rN_4}
\end{equation}
where we use the notation
$$
\rho_k^{(0)}=\rho_k.
$$

\section{Conclusion}

In future we will continue the study of the von~Neumann type equations and
their generalizations presented in the previous sections.
The integrals and the multi--soliton solutions will be considered.
An investigation of the hierarchies of symmetries and compatible flows for
these equations can lead to new hierarchies of integrable equations
\cite{MS,Sv,ASh,AMSh}.
The results will be of special interest in the case of integrable lattice
equations.
It should be also mentioned that discretizations of the lattice equations
attract large attention in recent years (see, e.g., \cite{LM} and references 
therein).

\section{Acknowledgements}
N. V. U. is grateful to Jan L. Cie\'sli\'nski and Marek Czachor for
hospitality.
This research was supported in part by the KBN grant 2 P03B 126 22
(J. L. C.), the KBN grant 5 P03B 040 20 (M. C.), 
Alexander von Humboldt foundation (M. C.) and Nokia--Poland (N. V. U.).

\section*{Appendix. Darboux Transformations}

Here we discuss briefly some particular cases of the Darboux transformation, 
which satisfy the conditions of Theorem in Sec.II.
Let the operator $D(\lambda)$ in (\ref{Tpsi}) be represented as given 
\begin{equation}
D(\lambda)={\bf1}+\frac{\nu-\mu}{\mu-\lambda}P,
\label{BDT}
\end{equation}
where
$$
P=\frac{\varphi\chi}{(\chi,\varphi)},
$$
$\chi$ is a solution of the Lax pair (\ref{Lax}) with parameter $\nu$:
$$
\left\{
\begin{array}{rcl}
-i\dot\chi&=&\chi A(\nu)\\
z_{\nu}\chi&=&\chi H(\nu)
\end{array}
\right.,
$$
$\varphi$ is a solution of the dual Lax pair with parameter $\mu$:
$$
\left\{
\begin{array}{rcl}
i\dot\varphi&=&A(\mu)\varphi\\
z_{\mu}\varphi&=&H(\mu)\varphi
\end{array}
\right.,
$$
$(\chi,\varphi)$ is a scalar product.
It is obvious that $P^2=P$ and
$$
-i\dot P=PA(\nu)P_{\perp}-P_{\perp}A(\mu)P
$$
($P_{\perp}={\bf1}-P$).
If coefficients of the transformed Lax pair (\ref{TLax}) are defined by
\begin{eqnarray}
A(\lambda)[1]&=&\sum_{k=0}^L\lambda^kB_k[1]+
\sum_{k=1}^M\frac{1}{\lambda^k}C_k[1],
\label{bTA}\\
H(\lambda)[1]&=&\sum_{k=0}^N\lambda^kH_k[1],
\nonumber
\end{eqnarray}
where
\begin{eqnarray}
B_k[1]&=&B_k+(\mu-\nu)\sum_{m=k+1}^L\left(\mu^{m-k-1}P_{\perp}B_mP-
\nu^{m-k-1}PB_mP_{\perp}\right),
\nonumber\\
C_k[1]&=&C_k-(\mu-\nu)\sum_{m=k}^M\left(\mu^{k-m-1}P_{\perp}C_mP-
\nu^{k-m-1}PC_mP_{\perp}\right),
\nonumber\\
H_k[1]&=&H_k+(\mu-\nu)\sum_{m=k+1}^N\left(\mu^{m-k-1}P_{\perp}H_mP-
\nu^{m-k-1}PH_mP_{\perp}\right),
\nonumber
\end{eqnarray}
then Eqs.\ (\ref{TLax}) are identically fulfilled.
This statement can be proved by direct computation.
Since
$$
D(\lambda)^{-1}={\bf1}+\frac{\mu-\nu}{\nu-\lambda}P,
$$
operators $D(\lambda)$ and $D(\lambda)^{-1}$ have poles in points $\mu$ and
$\nu$.
It is seen from Eq.\ (\ref{bTA}) that Eq.\ (\ref{mu_k}) are valid.

The formulas written above form the so-called binary Darboux transformation.
The corresponding Darboux transformation for an $n$--dimensional matrix case 
is produced from them if we assume in (\ref{BDT}) that
$$
P=\varphi(\chi\varphi)^{-1}\chi,
$$
where $\chi$ and $\varphi$ are respectively $m\times n$ and $n\times m$ matrix
solutions of direct and dual Lax pairs.
Some examples of Darboux transformations in infinite dimensional
case, which are suitable for integrable lattice equations, were given in
\cite{MatveevSalle}.
Very recently a new construction of the Darboux transformation in terms of
Clifford numbers was described in \cite{Ci2}.

\vfill
\eject
\end{document}